\documentstyle[epsf]{article}

\title{A family of complex potentials with real spectrum}

\author{
{\bf F. M. Fern\'andez}\\
 CEQUINOR (CONICET,UNLP)\\
Departamento de Qu\'imica,
Facultad de Ciencias Exactas\\
 Universidad Nacional de La Plata\\
 Calle 47 entre 1 y 115, Casilla de Correo 962\\
 1900 La Plata, Argentina,
\\
{\bf R. Guardiola}\thanks{E-mail: Rafael.Guardiola@uv.es}\\
Departamento de F\'{\i}sica  At\'omica y Nuclear\\
Universidad de Valencia\\
Avda. Dr. Moliner 50, 46100-Burjassot (Valencia), Spain,
\\
{\bf  J. Ros}
\\  Departamento de F\'{\i}sica Te\'orica and IFIC
\\   Universidad de Valencia\\
Avda. Dr. Moliner 50, 46100-Burjassot (Valencia), Spain
\\
and
\\
{\bf  M. Znojil}\thanks{E-mail: znojil@ujf.cas.cz}\\
 \'Ustav jadern\'e fyziky AV \v{C}R
\\  250 68 \v{R}e\v{z} u Prahy, Czech Republic
}
\date{\today}
 
\begin{document}
\maketitle

\newpage

\begin{abstract}
\noindent

We consider a two-parameter nonhermitean quantum-mechanical
Hamiltonian operator that is invariant under the combined
effects of parity  and time reversal transformations. Numerical
investigation shows that for some values of the potential
parameters the Hamiltonian operator supports real eigenvalues
and localized eigenfunctions. In contrast with other PT
symmetric models which require special integration paths in the
complex plane, our model is integrable along a line parallel to
the real axis.

\end{abstract}

\vspace{3cm} 


\newpage

\section{Introduction}
Recent analysis of the spectra of the family of 
Schr\"odinger operators with a
complex PT invariant potential \cite{bender1}
\begin{equation}
\label{xpown}
H = - \frac {d^2}{dx^2} - (ix)^\alpha
\end{equation}
has raised considerable interest on such class of operators
\cite{bender2}-\cite{milosh}. The motivation for the requirement
of Parity times Time Reversal (PT) (actually, complex conjugation)
 symmetry comes from a conjecture of D. Bessis relating it to the
existence of real energy bound states (in other words,
non-decaying resonances). From a practical point of view, PT symmetry is 
a way of selecting a specific Riemann sheet in the $(-\infty,0)$
cut complex coordinate plane, the cut being necessary to cope
with the $x=0$ branch point of the potential.

The loss of hermiticity of $H$ seems to imply complex
energies for localized eigenstates,  
with ${\rm Im}\ E_n\neq 0$ (i.e., levels with nonzero
widths), and consequently certain rate of decay of any initially
localized state.  
The loss of hermiticity may be mediated, e.g., by a spatial
asymmetry of the potential.   For example, it seems 
obvious that the cubic anharmonic Hamiltonian
\begin{equation}
H = - \frac{d^2}{dx^2} + x^2+g\,x^3
\label{AHO}
\end{equation}
allows particles to escape to infinity for any real
coupling $g \neq  0$.  

A deeper perturbative and Borel-summation
analysis of this problem proved 
such an oversimplified thumb rule wrong many years ago.  Calicetti et al
\cite{Calicetti} have shown that the spectrum of the cubic
anharmonic oscillator (\ref{AHO}) 
becomes {\em purely real} and positive at any purely imaginary
coupling $g = i\,g_I$ with $g_I > 0$. 
Moreover, the interpretation of the set of resonances 
of the Hamiltonian (\ref{AHO}) as analogous
to a set of bound states requires a suitable 
analytic continuation in $g$ and/or a careful deformation of the
integration path \cite{alvarez}.

The Schr\"odinger equation for the Hamiltonian operator  eq. (\ref{xpown}) 
cannot be integrated algebraically, and one has to resort to numerical
methods for its analysis. Cannata et al \cite{cannata1} have
found the way of generating complex Hamiltonians related to real
Hamiltonians, by applying a method invented by Darboux more than
one hundred years ago. Unfortunately, the resulting models have a
very complicated structure. 

In this paper we propose a family of problems characterized by
two parameters, see eq. (\ref{pot}) below, one of them ($\alpha$) 
playing a role analogous to the parameter $\alpha$ of eq. (\ref{xpown})
and the other ($\beta$) serving to {\em tune} the interaction
thus widening the richness of the spectrum. Our study of this family
will be numerical or, in other words, phenomenological, hoping
that it will help a further mathematical analysis.  One of
the main simplifications of our model is that the required complex
integration paths are lines parallel to the real axis, 
easy to implement and to interpret.

\section{The model}
We consider the two-parameter family of 
one-dimensional potentials 
\begin{equation}
\label{pot}
V_{\alpha \beta} = - (i \sinh x)^\alpha \cosh^\beta x,
\end{equation}
for arbitrary real values of $\alpha$ and $\beta$. 
These functions have in general a branch point at $x=0$ and we 
select such a branch that the real part of the potential
is {\em symmetric} and the imaginary part {\em antisymmetric}
with respect to the origin. Specifically, the potential will be
defined by the two equations
\begin{eqnarray*}
x>0  & V(x)= & e^{i \pi (2+\alpha)/2} \sinh^\alpha x \, 
\cosh^\beta x \\
x<0  & V(x)= & e^{-i \pi (2+\alpha)/2} \sinh^\alpha |x| \, 
\cosh^\beta x 
\end{eqnarray*}
thus having an invariant Hamiltonian under the PT
transformation, i.e., parity transformation and complex conjugation.
This requires to cut the complex $x$-plane from $x=0$ up to $x=-\infty$, 
and to consider  the
relevant negative $x$ values below the cut, i.e. with a small
negative imaginary part or a phase $-\pi$.

\begin{table}[htb]
\caption{Variability of our PT symmetric potential with
 parameter $\alpha$, for $x>0$ and any real $\beta$}
\label{tab.pot}

\vspace{0.5cm}
\begin{tabular}{|lll|}
\hline
\hline
$\alpha$ & Real part & Imaginary part \\
\hline
0       & $V_R<0$      & $V_I=0$ \\
$0 < \alpha < 1 $& $V_R < 0$ & $V_I < 0 $\\
$\alpha = 1 $& $V_R = 0$ & $V_I < 0 $\\
$1 < \alpha < 2 $& $V_R > 0$ & $V_I < 0 $\\
$\alpha =2  $& $V_R > 0$ & $V_I = 0 $\\
$2 < \alpha < 3 $& $V_R > 0$ & $V_I > 0 $\\
$ \alpha =3 $& $V_R = 0$ & $V_I > 0 $\\
$3 < \alpha < 4 $& $V_R < 0$ & $V_I > 0 $\\
$ \alpha = 4 $& $V_R < 0$ & $V_I = 0 $\\
\hline
\hline
\end{tabular}
\end{table}

The characteristic values of the potential for different values
of the parameter $\alpha$ are shown in Table \ref{tab.pot}, where
we only see the region $\alpha \in [0,4]$, because the same
structure is repeated for larger values of $\alpha$ with period 4. 
In all cases,
both the real and imaginary parts of the potential tend to
either $+\infty$ or $-\infty$ at long distances, 
 when $x \rightarrow \pm \infty$. The value of $\beta$ does not
affect the main
features of the potential, except for the fact that negative values of 
$\beta$ may give rise to
non-confining potentials. We exclude such cases from present 
study and always consider $\alpha+\beta >0 $.

The interesting point about PT symmetric Hamiltonians 
\begin{equation}
\label{ham}
H = - \frac{d^2}{dx^2} + V_{\alpha \beta}(x)
\end{equation}
is that they may have {\em localized} solutions, corresponding to real
eigenvalues, which may be interpreted either as bound states or as 
zero width resonances. 

The special case $\alpha=2$ corresponds to a positive and confining
potential, with an infinite number of bound states. It will be a
suitable reference point for calculations corresponding to other values of
$\alpha$. We will move towards $\alpha>2$ and $\alpha<2$ starting
from a given eigenvalue of the case $\alpha=2$ to show the
evolution of the real-energy eigenstates.

Analogously, one may take any other reference  value of
$\alpha = \alpha_R$ which produces a real and positive confining potential,
like $\alpha_R=6$, $\alpha_R=10$, and so on. 
Selecting a different value of $\alpha_R$ means to consider 
a different eigenvalue problem, which will be  labeled by that
particular reference value $\alpha_R$. Here we will concentrate on 
the case $\alpha_R=2$ mentioned above, and rise considerations 
also for other values of $\alpha_R$.

This family of potentials is similar to the one-parameter
family $V(x)=-(i x)^\alpha$ recently considered by Bender and
Boettcher \cite{bender1}, 
but the exponential growth of our potentials at long
distances simplifies the analysis of their properties.

\section{Paths in the complex plane}
By carrying out the integration of the Schr\"odinger equation
\begin{equation}
\label{SE}
- \frac {d^2 \Psi(x)}{d x^2} + V_{\alpha \beta} \Psi(x) = E \Psi(x)
\end{equation}
along the real axis (see below for more details), 
for values of $\alpha\in [0,4[$, one
observes that there are solutions for real values of the
energy $E$, smoothly connected with the 
solutions of the real potential with $\alpha=2$.
The wave functions are  complex, and  may be chosen 
to have a symmetric real part and an antisymmetric
imaginary part. The last statement is a consequence of the PT
invariance, which requires that $\Psi^*(-x)$ is a solution of
the Schr\"odinger equation if $\Psi(x)$ is a solution for a {\em
real} eigenvalue $E$. On choosing an appropriate phase factor
one may then have the mentioned symmetries for the localized
wave functions.

\begin{figure}[htb]
\caption{Integration of the Schr\"odinger equation along the real $x$-axis,
for the model potential $V_{\alpha \beta}$. Only the ground-state
energy is shown. 
The two branches correspond to solutions smoothly connected with
the eigenvalues for $\alpha_R=2$ (left curve) and $\alpha_R=6$
(right curve).
The dashed line depicts the expected 
eigenvalues after an appropriate analytic
continuation into the complex $x$-plane, for the solutions smoothly
connected with $\alpha_R=2$.}
\label{fig:1}
\centerline{
\epsfxsize=10cm \epsfbox{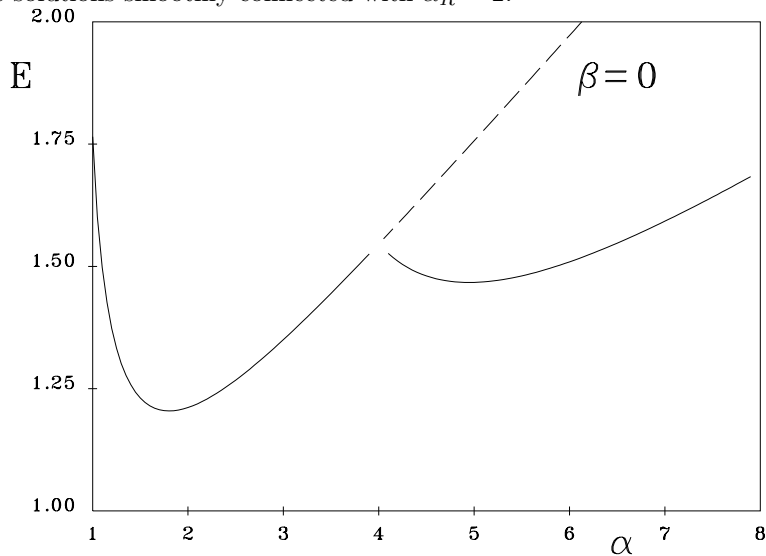}}
\end{figure}

As discussed above, the point $\alpha=2$ is an exception,
because the Hamiltonian is separately parity and time-reversal
invariant. Therefore one can choose the wave functions to be
real, and either symmetric or antisymmetric with respect to $x=0$.

At $\alpha=4$ the potential is real and everywhere negative;
consequently there are no eigenstates with real eigenvalues.
Real eigenvalues appear again for $\alpha \in ]4,8[$, which 
are smoothly connected with the case $\alpha=6$. The same
pattern is repeated at every $\alpha=2+4N$, for positive
integer values of N, as we increase $\alpha$.

These facts are shown in Fig. \ref{fig:1}, for the case $\beta=0$.
There, we clearly see the existence of two (in general, many)
different problems, one centered at $\alpha_R=2$ and the other
centered around $\alpha_R=6$. The purpose of this section is to
find a path in the complex $x$-plane to obtain lines like the
dashed one, which represents the continuation of the eigenvalue
beyond the limit $\alpha=4$.

The first question to consider is whether there are 
confined solutions. To this end we take into account 
the limit $x \rightarrow +\infty$ in the Schr\"odinger equation, 
where the potential is dominated by the exponential part
$$
V_{\alpha \beta}(x) 
\rightarrow \exp[i \pi(2+\alpha)/2] \exp[(\alpha+\beta) x] 
/2^{\alpha+\beta}.
$$
One proceeds as in the WKB method assuming a general solution 
of the form $\Psi(x) = \exp[G(x)]$. The leading order of the
asymptotic expansion for $G(x)$ is 
\begin{equation}
\label{phase}
G(x) \rightarrow 
\pm e^{i \pi(2+\alpha)/4} \frac{2 e^{(\alpha+\beta) x/2}}
{2^{(\alpha+\beta)/2} (\alpha+\beta)},
\end{equation}
where the plus and minus signs come from a square root which
appears in the differential equation for $G(x)$.

Except for the particular cases $\alpha=4N$, N=0, 1, \dots, there
appears to exist a solution such that the real part of $G(x)$ is negative
and its magnitude increases exponentially, suggesting a discrete 
set of eigenvalues with localized solutions. 
As mentioned above, we require $\alpha+\beta>0$ in order to 
have asymptotically vanishing solutions at long distances.

The only general statement to be drawn from the above
asymptotic limit is that, as far as the phase $\pi (2+\alpha)/4$
is different from a half integer multiple of $\pi$, there are
two possible asymptotic solutions, one growing and the other one
decreasing at long distances; therefore, one one may expect to 
find one or more values of $E$ that select the asymptotically
vanishing solutions corresponding to localized states.

\subsection{The $\alpha_R=2$ family}
When $\alpha=2$ the phase factor in eq. (\ref{phase}) is 
$\exp(i\pi)$, and the required solution is the one with the plus sign
in front of it (remember that $\Psi=\exp [G(x)]$). For 
$\alpha=2+\delta$ the phase factor changes to $\pi + \delta \pi/4$, and
the solution behaves asymptotically as
$$
\Psi(x) \rightarrow \exp\left[ [ -\cos (\delta \pi/4) -i
\sin(\delta \pi/4) ] e^{(\alpha+\beta) x/2}
\right] ,
$$
when $x\rightarrow \infty$; i.e., the exponentially decreasing 
part subsists as far as $|\delta| < 2$. The farther we move 
from $\alpha=2$ the slower the 
exponentially decreasing part approaches to zero, indicating that the
asymptotic regime will be reached at much larger distances.

There is a way of both moving {\em Asymptotia} closer and extending the
integration beyond $\alpha=4$, which consists in adding to the
integration variable an imaginary component $i y$ which should
be negative to be consistent with the potential branch required 
by the PT symmetry. After the replacement $x \leftarrow x+iy$
in eq. (\ref{phase}) the phase angle 
of the auxiliary function $G$ is  changed to
\begin{equation}
\label{angle}
\theta =
\frac{\pi (\alpha+2)}{4} + \frac{y(\alpha+\beta)}{2}.
\end{equation}
The line corresponding to $\theta=\pi$ labels the path of
fastest decrease of the exponential
(the {\em anti-Stokes} lines of Ref. \cite{bender1}), and the 
boundaries of the region of convergence result from the solutions 
of $\theta=\pi \pm \pi/2$.

\begin{figure}[htb]
\caption{The continuous line represents the preferred
value of the imaginary part $y$ added to the coordinate $x$,
for each value of $\alpha$ and for the family of eigenstates
smoothly connected with $\alpha_R=2$. The two dashed lines are the
boundaries of the acceptable values of $y$.}
\label{path2}
\centerline
{ \epsfxsize=10cm \epsfbox{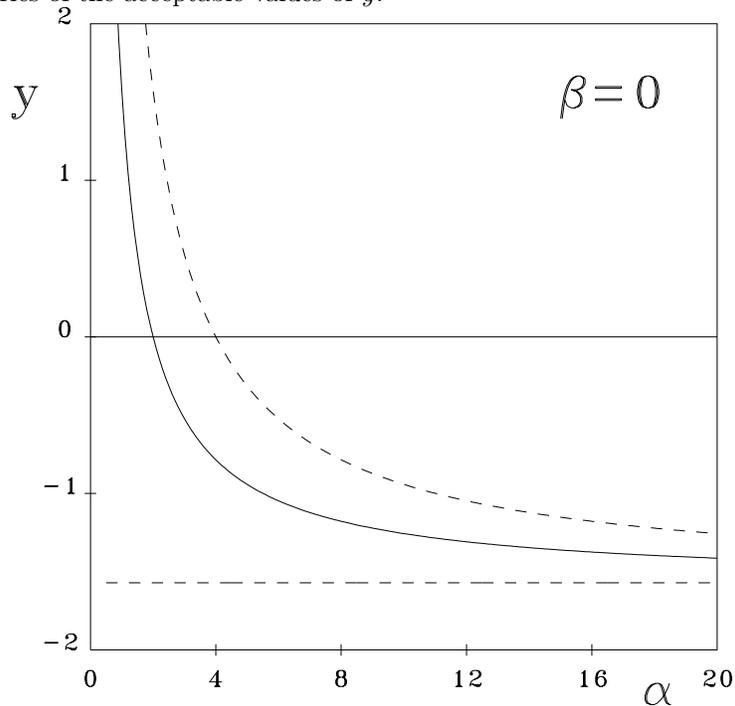}}
\end{figure}

The optimal path corresponds to a value of $y$ given by
\begin{equation}
\label{optimal}
y = \frac{(2-\alpha)\pi}{2(\alpha+\beta)} 
\end{equation}
and remains negative for any value of $\alpha>2$, fulfilling the
requirement for PT symmetry. The integration, however, may be
performed for any value of $y$ in the range
\begin{eqnarray*}
y_+ &=& {\displaystyle
\frac{(4-\alpha) \pi}{2(\alpha+\beta)}} \\
y_- &=& {\displaystyle
- \frac{\alpha \pi}{2(\alpha+\beta)}}
\end{eqnarray*}
so that, for $\alpha<2$ the integration should be carried out along
the real $x$-axis to satisfy PT symmetry. 
The optimal and boundary values of $y$ are plotted in Fig. \ref{path2}.
We see that without violating PT symmetry, the integration may
be carried out along the real axis up to $\alpha=4$,

Here our model differs significantly from that considered by
Bender and Boettcher \cite{bender1}. In their model, eq. (\ref{xpown}),
 the integration
must be carried out along two symmetric sectors, one in the lower-right
complex $x$-plane, and the other symmetric with respect to the
imaginary axis. The optimal line is given by $x \exp(i\theta)$,
where $\theta= -(\alpha-2)/(\alpha+2) ( \pi /2) $, in such a
way that it tends to coincide with the negative imaginary axis
for large values of $\alpha$. In our case the displacement $y$ 
remains bounded in the large-$\alpha$ limit, having the value
$-\pi/2$.

\subsection{The families for $\alpha_R=6$ and beyond}
We can carry out an analogous study for the family of
potentials connected to $\alpha=6, 10, \dots$ For the case
$\alpha=6$ one obtains the optimal displacement 
$$
y = \frac{(6-\alpha)\pi}{2(\alpha+\beta)} 
$$
and the lower limit 
$$
y_- =  \frac{(4-\alpha)\pi}{2(\alpha+\beta)} 
$$
which is suitable for the integration when $\alpha>4$.

To extend the integration to $\alpha<4$ we choose an 
alternative path given by the solution with minus sign in eq. 
(\ref{phase}), leading to an optimal displacement 
$y=-\pi(\alpha+6)/2(\alpha+\beta)$.

After this discussion we clearly understand the peculiar behavior
of the lowest real eigenvalue shown in Fig. \ref{fig:1}. The
integration having been carried out along the real axis, the
lowest eigenvalues jump from family to family when crossing the
points where the potential is purely imaginary. Actually, the
way of obtaining the eigenvalue by requiring that only the
normalizable component survives selects the plus or
minus sign in eq. \ref{phase}. In this way, in addition to the
lack of continuity in the eigenvalues, one may
also observe a sudden jump in the phase of the wave function 
(ie, the sign of the imaginary part near $0^+$) when crossing
each special case mentioned above.

\subsection{The role of the shift $y$}
It is not difficult to understand the role of the change of variable
$x \rightarrow x+iy$ if we just carry it out explicitly in eq. (\ref{SE}).
The potential-energy function in the resulting Schr\"{o}dinger 
equation reads 
$$ V_{\rm eff}(x) = V(x+iy),$$
so that the new potential with the above mentioned values of 
$y$ has a dominant confining real part and a much smaller imaginary part.

\begin{figure}[htb]
\caption{Real (left) and imaginary (right) parts of the
effective potential obtained by shifting the real variable $x$
to the lower part of the complex plane with the optimal $y$ value.
The potentials correspond to the phase selection appropriate for
 the family smoothly connected with $\alpha=2$. The calculations
were performed with $\beta=0$ for three values of $\alpha=2, 3$
and $4$, which label the corresponding curve}
\label{veff}
\centerline{
\epsfxsize=10cm \epsfbox{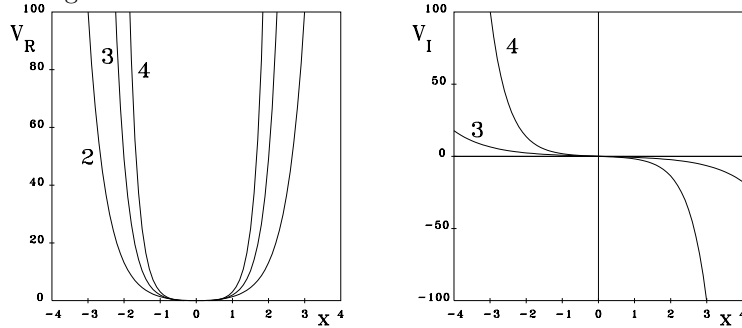}
}
\end{figure}

The actual effect of the transformation is shown in Fig. \ref{veff}
for three values of $\alpha$. Particularly impressive is the
case $\alpha=4$ which originally was a real and negative
potential, and after the transformation exhibits a dominant real 
confining component. The transformation of the potential guarantees 
the connection through the special points $\alpha=4N$.

\section{Numerical integration}
The numerical calculation of the eigenvalues of the Schr\"odinger
equation (\ref{SE}) with a complex potential and along a complex
path, is as simple as in the case of real potentials along a real
path.

We have chosen the simplest algorithm which starts by selecting
two extreme points, $X_{\rm min}$ and $X_{\rm max}$, at which
the wave function is assumed to vanish, and discretizing this
interval with a uniform integration step $h$, defined in terms
of the number of points $N$ as
$$ h = \frac{X_{\rm max}- X_{\rm min}}{N+1} .
$$
In order to preserve the PT invariance in the discretization, it
is necessary to take $X_{\rm min}=-X_{\rm max}$. An integer counter 
$k$ labels the mesh points as $x_k = X_{\rm min}+k h$.
Approximating the second derivative by the second differences operator,
$$
\frac{d^2 \Psi_k}{dx^2} \simeq \frac {\Psi_{k+1} - 2 \Psi_{k} +
\Psi_{k-1} }{h^2},
$$
the continuous eigenvalue problem becomes a discrete one 
given by a {\em symmetric} and {\em tridiagonal} matrix of dimension $N$ 
and matrix elements
\begin{eqnarray}
\label{ME}
\nonumber
H_{ii} & = & \frac{2}{h^2} +V_k \\
H_{i,i+1} & = & -\frac{1}{h^2}.
\end{eqnarray}
It is understood that $V_k = 
V(X_{\rm min}+k h + i y)$ in the equation above, 
where $y$ is the appropriate 
imaginary shift already described earlier.

The tridiagonal matrix is symmetric, but, contrarily to the case
of a real potential, it is not hermitean. The roots of the
determinant 
$$
 D_N(E)=\det [ H - I E] =0
$$
give the eigenvalues approximately. The calculation being 
greatly facilitated by the three-point recurrence relation
\begin{equation}
\label{sequence}
 D_n(E) = D_{n-1}(E) ( H_{nn}-E) - H_{n,n-1}^2 D_{n-2},
\ \ n=2,3,\dots ,N
\end{equation}
with the starting conditions $D_0$=1 and $D_1=H_{11}-E$.
This recurrence relation exhibits the same structure as in the case
of a real potential, but not the same properties. In the case of
a real potential every term of the sequence is real, and the set
constitutes a {\em Sturm sequence} \cite{wilkinson,golub}. This
property allows one to devise a simple search algorithm, combining
the counting of sign changes of the sequence with the bisection
method to determine efficiently the eigenvalues.

This scheme is not appropriate for our case because the
potential is complex. 
However, an important feature of our approach is that PT
symmetric potentials give rise to determinants which are
polynomial functions $E$ with real coefficients. The proof is quite
simple. First one notices that the diagonal matrix
elements satisfy the rule
$$
H_{ii} = H_{N-i+1,N-i+1}^*,
$$
consequence of the PT symmetry. So, the time reversal operation
T, which corresponds to changing each element of the diagonal by
its complex conjugate, is equivalent to applying the parity P
operation, which corresponds to changing every element of the
diagonal by its symmetric with respect to the centre of the
diagonal (this operation is carried out by a similarity
transformation with an orthogonal matrix having all elements
equal to zero with the exception of 
the counter-diagonal elements which are unity).

In conclusion, the determinant of $H-EI$ is real for
real values of E. This is true only if $X_{\rm min}=-X_{\rm max}$,
in such a way that only the determinant $D_N$ is real, but not
the terms of the sequence $D_n$ with $n< N$.
This feature allows the use of the robust bisection method to determine the
eigenvalues.

In almost all cases, we have set the value of $\alpha$ and then determined 
the corresponding eigenvalue. However, in the neighborhoods of
the points where two real eigenvalues collapse into a pair
of complex conjugate eigenvalues, it is more
convenient to determine the value of $\alpha$ for a given
eigenvalue. In any case, the method is simple and robust.

\section{The mutual interplay of $\alpha$, $\beta$ and energies}
Having arrived at the proper way of extending our calculations
beyond the special points $\alpha=4N$, our next step is the
recomputation of Fig. \ref{fig:1} including some excited levels.

\begin{figure}[htb]
\caption{Spectrum of the PT invariant Hamiltonian with the
potential $V_{\alpha \beta}$ and $\beta=0$, showing several
bound states corresponding to two families, one connected with
$\alpha=2$ (continuous line) and the other connected with $\alpha=6$
(dashed lines)}
\label{twofam}
\centerline
{
\epsfysize=6cm \epsfbox{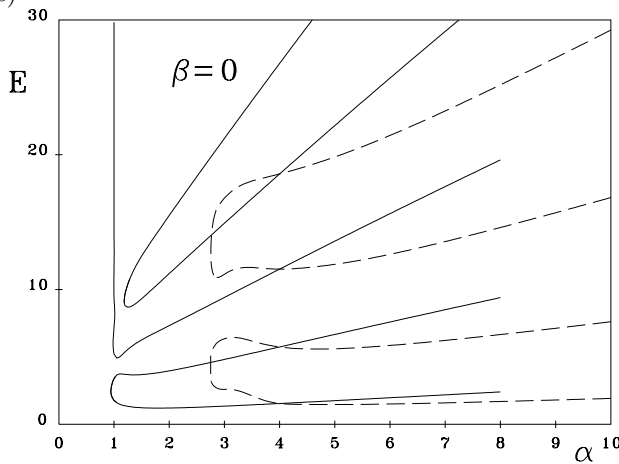}
}
\end{figure}

 Fig. \ref{twofam} shows results for several 
levels and for the two sets smoothly connected with 
$\alpha=2$ and $\alpha=6$. The main feature of this figure is
that there is a one-to-one correspondence between an eigenvalue
with $\alpha$ close to $2$ and an eigenvalue of the real
confining potential with $\alpha=2$. The same situation takes
place at $\alpha=6$, and it is easily proved by means of
perturbation theory for $\alpha$ close to $2, 6,\dots$

Our numerical calculations suggest that there will be real
eigenvalues within each family with $\alpha$ greater than the
reference value $\alpha_R=2$, and that for smaller values of
$\alpha$ the real eigenvalues merge into pairs of complex
conjugate values of $E$, until reaching the vicinity of $\alpha=1$
where once again real eigenvalues are allowed.

\begin{figure}[hbt]
\caption{Enlarged view of the spectrum for $\beta=0$ near $\alpha=1$.
}
\label{enlarg}
\centerline
{
\epsfysize=6cm \epsfbox{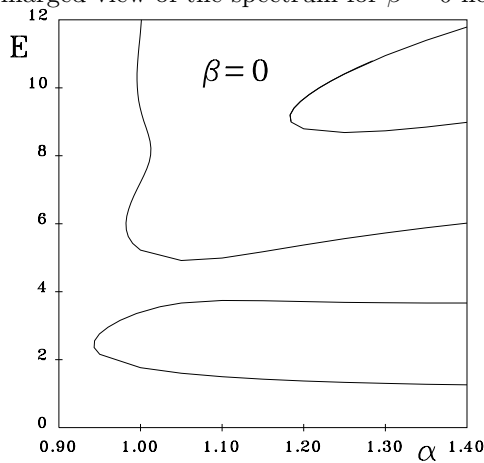}
}
\end{figure}

Figure \ref{enlarg} shows in detail the special characteristics
of the levels near $\alpha=1$ for $\beta=0$. In particular, this
 figure illustrates the simultaneous jump of the fourth and fifth
levels into the complex plane, near $\alpha=1.15$, and their 
simultaneous return to the real axis when $\alpha$ is slightly
greater than unity. The same pattern seems to happen also for higher
levels. Obviously, as far as the characteristic polynomial is
real, the transition from real to complex values must be in pairs.
We have not observed a similar phenomenon in the case of the set
connected with $\alpha=6$, but it may well happen for levels of
energy higher than those shown in Fig. \ref{twofam}.

\begin{figure}[htb]
\caption{Several eigenvalues computed at $\beta=0.5$,
$\beta=0.25$, $\beta=0$ and $\beta=-0.25$}
\label{mapa}
\centerline
{
\epsfysize=10cm \epsfbox{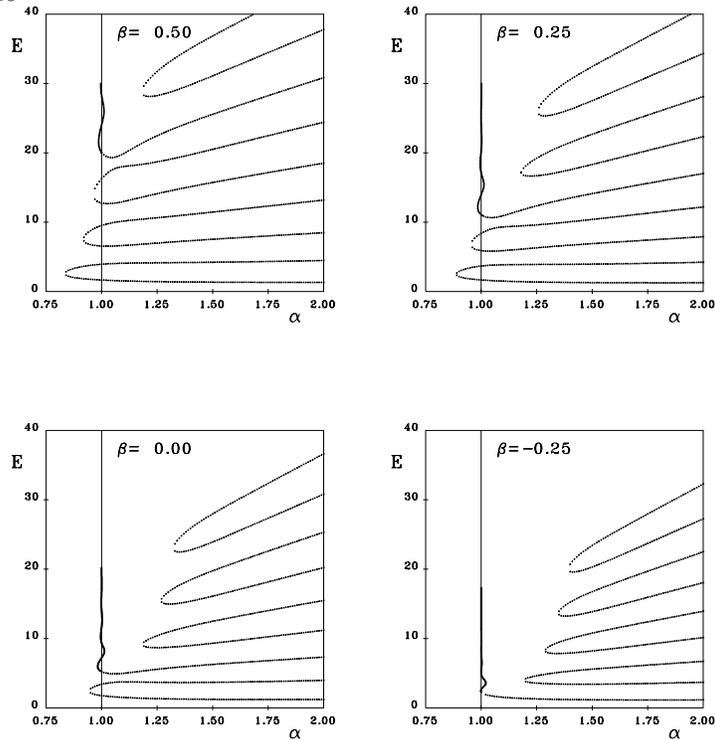}
}
\end{figure}

Up to now we have concentrated on calculations with $\beta=0$.
Fig. \ref{mapa} illustrates the role of the parameter $\beta$. 
In addition to the case $\beta=0$, this figure displays
also the lowest levels for several values of
$\beta$, both positive and negative. The interesting role of
$\beta$ is to switch the {\em special} level, i.e., the level which
ultimately will move around $\alpha=1$. With its help one may
choose this special level to be the first one, ($\beta=-0.25$), 
the third one ($\beta=0$), the fifth one ($\beta=0.25$) and 
so on. Because of the special way the levels form the pairs, 
one should not be surprised that even levels cannot become 
{\em special} in the above mentioned sense.

\section{Quasi-algebraic study}
As in our previous paper on the subject, we have also
supplemented the numerical calculation with the Riccati-Pad\'{e}
method (RPM).  When $\alpha=2$ and $\beta=0$ the potential-energy
function is parity-invariant and the RPM leads to just one
Hankel determinant from which one obtains the eigenvalues
\cite{RPMSIM}. The calculation is straightforward and the rate of
convergence sufficiently great as shown in Table \ref{RPM1}.

\begin{table}[htb]
\caption{The RPM ground-state energy for $\beta=0$ and $\alpha=2$
in terms of the dimension of the Pad\'e determinant}
\label{RPM1}

\vspace{0.5cm}
\begin{tabular}{|ll|}
\hline
hline
$D$  &  RPM root\\
\hline
2   &  1.213616523 \\
3   &  1.211409311 \\
4   &  1.211411109 \\
5   &  1.2114109830 \\
6   &  1.211410984169 \\
7   &  1.2114109841755 \\
8   &  1.21141098417527 \\
\hline
\hline
\end{tabular}
\end{table}

\begin{table}[hbt]
\caption{The RPM ground-state energies for 
the non-hermitean cases  $\alpha=1$ and $\alpha=3$.
In both cases is $\beta=0$.}
\label{RPM2}

\vspace{0.5cm}
\begin{tabular}{|lllll|}
\hline
 &
 $\alpha=1$    &  $\alpha=1$&
 $\alpha=3$    &  
 $\alpha=3$    \\  
\hline
$D$ 
&$ E$ & $i \Psi'(0)/\Psi(0)$ &
$ E$ & $i \Psi'(0)/\Psi(0)$ \\
\hline
2   & 1.655005966 &   1.033573034 &
- -&-\\
3   & 1.765033153 &  1.095023981 &
    1.385656774 & -0.5049697062 \\
4   & 1.765157398 &   1.095137449 
  & 1.349869536 & -0.4769952880 \\
5   & 1.765157246 &   1.095137384 
  & 1.350149473 & -0.4771536171 \\
6 & - & -
  & 1.350140759 & -0.4771520606 \\
Numeric & 1.76515725 & 1.09513737 &
 1.350140990 & -0.47715200\\
\hline
\hline
\end{tabular}
\end{table}

\begin{table}[htb]
\caption{The RPM ground-state energies for 
the non-hermitean cases  $\alpha=1$ and $\alpha=3$ with the
complex rescaled Hamiltonian.
In both cases is $\beta=0$.}
\label{RPM3}

\vspace{0.5cm}
\begin{tabular}{|lll|}
\hline
$D$   &  $ E(\alpha=1)$ &  $ E(\alpha=3)$ \\
\hline
 2 & 1.765248635 &  2.60904086409 \\ 
3 & 1.765157328 & 2.59510727493   \\ 
4 & 1.765157255 & 2.59524841637   \\ 
5 & 1.76515725525231 & 2.59524599823   \\ 
6 & 1.76515725525336 & 2.59524605087   \\ 
7 & 1.7651572552533587 & 2.59524605034 \\ 
8 & 1.76515725525335874 & - \\
\hline
\end{tabular}
\end{table}

For nonhermitean cases we change the coordinate according to
$x=iq$ so that the Hamiltonian operator becomes 
$$-H= - \frac {d^2}{dq^2}+[- \sin (q)]^ {\alpha} \cos(q)^ {\beta} $$
and we can apply the RPM as in the case of a real
Schr\"{o}dinger equation. If the potential energy is an even
function of $q$ we apply the method just indicated; if it is not
then the RPM leads to two Hankel determinants \cite{RPMASIM}
from which we obtain both $E$ and $-i \Psi'(0)/\Psi(0)$. Table
\ref{RPM2} shows results for $\alpha=1$ and $\alpha=3$ in
excellent agreement with the numerical integration discussed
above. 

We have carried out the RPM calculations algebraically by means
of Maple, resorting to a numerical approach just at the end
in order to obtain the roots of the Hankel determinants
\cite{RPMSIM,RPMASIM}. For this reason the requirement of
computer memory is considerable in the case of a nonsymmetric
potential-energy function, and we cannot handle determinants of
the same dimension as in the symmetric case. For $\beta=0$ we
have tried to overcome this problem by means of the change of
coordinate $x=i(u+ \pi/2)$ \cite{Moiseyev} and applying RPM for
symmetric potential-energy functions to the resulting
Hamiltonian operator 
$$-H=-\frac {d^2}{du^2} + [-\cos \, u ]^{\alpha}.$$
Table \ref{RPM3} shows results for $\alpha=1$
and $\alpha=3$. In the former case the result is identical
(though more accurate) to the one shown in Table \ref{RPM2};
however, in the latter case we do not obtain the lowest
eigenvalue but the first excited one. We have not yet being able
to justify this anomalous behavior of the RPM when $\alpha=3$.

\section*{Acknowledgments}

RG and JR are supported by DGES under contract Nb. PB97--1139.
MZ acknowledges the
financial support via grants A 104 8602 (GA AV \v{C}R) and
202/96/0218 (GA \v{C}R).

\end{document}